\documentclass[aps,prd,twocolumn,superscriptaddress,preprintnumbers,floatfix,nofootinbib,notitlepage,showkeys,showpacs]{revtex4-2}
\usepackage[utf8]{inputenc}
\usepackage{epsfig,amsfonts,amsthm}
\usepackage{epstopdf}
\usepackage{amsmath,amssymb}
\usepackage{latexsym}
\usepackage{bbm}
\usepackage{array}
\usepackage{comment}
\usepackage{hyperref}

\usepackage{color}
\usepackage{pb-diagram}
\usepackage{slashed}
\usepackage{cancel}
\newcommand{\be}{\begin{equation}}
\newcommand{\ee}{\end{equation}}
\newcommand{\bea}{\begin{eqnarray}}
\newcommand{\eea}{\end{eqnarray}}

\usepackage{color}
\definecolor{myorange}{rgb}{1,0.5,0}

\begin{document}
\bibliographystyle{OurBibTeX}


\title{Daily and annual modulation rate of low mass dark matter in silicon detectors}

\author{Abolfazl Dinmohammadi}
\email{dinmohammadi@znu.ac.ir}
\affiliation{Department of Physics, Faculty of Science, University of Zanjan,  P.O. Box 45195-313, Zanjan, Iran}

\author{Matti Heikinheimo}
\email{matti.heikinheimo@helsinki.fi}
\affiliation{Department of Physics, University of Helsinki, 
                      P.O.Box 64, FI-00014 University of Helsinki, Finland}
\affiliation{Helsinki Institute of Physics, 
                      P.O.Box 64, FI-00014 University of Helsinki, Finland}

\author{Nader Mirabolfathi}
\email{mirabolfathi@physics.tamu.edu}
\affiliation{Department of Physics and Astronomy, and the Mitchell Institute for Fundamental Physics and
Astronomy, Texas A$\&$M University, College Station, TX 77843, USA}

\author{Kai Nordlund}
\email{kai.nordlund@helsinki.fi}
\affiliation{Department of Physics, University of Helsinki, 
                      P.O.Box 64, FI-00014 University of Helsinki, Finland}
\affiliation{Helsinki Institute of Physics, 
                      P.O.Box 64, FI-00014 University of Helsinki, Finland}

\author{Hossein Safari}
\email{safari@znu.ac.ir}
\affiliation{Department of Physics, Faculty of Science, University of Zanjan, P.O. Box 45195-313, Zanjan, Iran}
 
\author{Sebastian Sassi}
\email{sebastian.k.sassi@helsinki.fi}
\affiliation{Department of Physics, University of Helsinki, 
	P.O.Box 64, FI-00014 University of Helsinki, Finland}
\affiliation{Helsinki Institute of Physics, 
	P.O.Box 64, FI-00014 University of Helsinki, Finland}
 
\author{Kimmo Tuominen}
\email{kimmo.i.tuominen@helsinki.fi}
\affiliation{Department of Physics, University of Helsinki, 
                      P.O.Box 64, FI-00014 University of Helsinki, Finland}
\affiliation{Helsinki Institute of Physics, 
                      P.O.Box 64, FI-00014 University of Helsinki, Finland}

\begin{abstract}
\noindent
{Low threshold detectors with single-electron excitation sensitivity 
to nuclear recoil events in solid-state detectors are also sensitive to the crystalline structure of the target and, therefore, to the recoil direction via the anisotropic energy threshold for defect creation in the detector material. We investigate this effect and the resulting daily and annual modulation of the observable event rate for dark matter mass range from 0.2 to 5 GeV/c$^{2}$ in a silicon detector. We show that the directional dependence of the threshold energy and the motion of the laboratory result in modulation of the event rate which can be utilized to enhance the sensitivity of the experiment. We demonstrate that the spin-independent interaction rate in silicon is significant for both high and low dark matter masses. For low-mass dark matter, we show that the average interaction rate in silicon is larger than germanium, making silicon an important target for identifying dark matter from backgrounds. We find 8 and 12 hours periodicity in the time series of event rates for silicon detector due to the 45-degree symmetry in the silicon crystal structure.
}
 \end{abstract}
\preprint{HIP-2023-2/TH}
\maketitle
\section{Introduction}
Observations of large-scale cosmic phenomena, galaxy clusters,  the matter power spectrum and cosmic microwave background radiation are evidence for the existence of non-baryonic matter that constitutes about 85\% of the total matter content of the Universe~\cite{Kadribasic:2017, Ade:2014}. Non-baryonic matter, known as dark matter, is one of the most fundamental topics in cosmology, astronomy, and high-energy physics~\cite{Cushman:2013,aghanim:2020,bertone:2005}. Most dark matter laboratories are designed and implemented based on direct and indirect detection approaches. 
The indirect detection approach corresponds to the annihilation or decay of dark matter particles~\cite{Jungman:1996}. In contrast, direct detection is via the elastic collision of a dark matter particle with the nuclei of atoms~\cite{Gaitskell:2004} and generates typical recoil energy of 
${\mathcal O({\mathrm{ keV}})}$~\cite{Goodman:1985,OHare:2015}. This energy operates to create an electron-hole pair inside the detector. In recent  decades, many studies for the direct search of dark matter were employed, such as XENON10~\cite{Xe10}, XENON100~\cite{Xe100:2012}, SIMPLE~\cite{Girard:2012}, CoGeNT~\cite{Aalseth:2013} and DAMA/LIBRA~\cite{Bernabei:2010, Bernabei:2018}.
In most direct dark matter experiments, a dark matter particle has been considered with a mass of ${\mathcal O(10-100)}$ GeV/c$^{2}$ 
~\cite{Cushman:2013}. Recently, with the development of convincing theoretical models, dark matter masses less than 10 GeV/c$^{2}$ have also been introduced~\cite{Kadribasic:2017, Cushman:2013, Romani:2018, Crisler:2018}.

In many experiments, the target material is either low-pressure gas or scintillating liquids. These experiments typically feature a rather high threshold energy for detecting dark matter based on nuclear  recoil~\cite{Kadribasic:2017, Mayet:2016}. Also, in a gaseous detector, the large volume of gas required is a deterrent. Due to the low energy of nuclear recoils for dark matter with low mass, high-mass detectors with low detection thresholds are more desirable \cite{Agnese:2014, Mirabolfathi:2017,Iyer:2021}. Semiconductor targets, such as silicon or germanium, can be utilized as cryogenic solid-state ionization detectors, where a sensitivity to single electron events has been demonstrated~\cite{Kadribasic:2017, Mirabolfathi:2017, Agnese:2018,Heikinheimo:2022}. When a particle colliding with a nucleus of the detector material transfers enough energy to the nucleus, it causes a defect in the configuration of the crystal lattice~\cite{Kadribasic:2017}. The effective threshold energy in generating the defect depends on the recoil angle in silicon and germanium crystals. Holmstrom et al.~\cite{Holms:2008, Holms:2010} obtained the threshold energy required to create a defect in germanium and silicon crystals for different directions using molecular dynamics simulations.  It has been argued~\cite{Heikinheimo:2019,Kadribasic:2017} that this threshold displacement energy is equivalent to the ionization threshold, which implies that the ionization signal will also be sensitive to the recoil direction.
Like most spiral galaxies, the current models assume that the Milky Way is immersed in a halo of dark matter. Inside the halo, the solar system moves toward the constellation Cygnus~\cite{OHare:2015}, and the direction of motion of dark matter particles in the lab frame is opposite to this. The Earth's rotation and motion around the Sun cause the laboratory frame velocity of the dark matter particles to change over time. Due to the anisotropy of the ionization threshold, this results in daily and annual modulation effects in the dark matter interaction rate~\cite{Heikinheimo:2019,sassi:2021,Coskuner:2019,Trickle:2019,Coskuner:2022}.

In this paper, we study the daily and annual rate of dark matter interaction with silicon nuclei, taking into account the energy threshold and the direction of the recoil induced by the collision with a dark matter particle. We use threshold energy data for silicon obtained via molecular dynamics simulations. We investigate the modulation of the ionization rate for dark matter interactions due to the directional dependence of the threshold energy and the motion of the laboratory frame with respect to the galactic rest frame. 

The layout of this paper is as follows: in Section~\ref{sec2} we calculate the directional event rate of a dark matter in the presence of a direction-dependent energy threshold, and in Section~\ref{sec3} we present the results. 

\section{Dark matter rate }\label{sec2}
To calculate the interaction rate of dark matter with the nuclei of the detecting material, we need the distribution function of the dark matter velocity in the galactic halo. We consider the Maxwell-Boltzmann distribution~\cite{lewin:1996, sassi:2021} for the velocity of dark matter in the halo as follows:
\begin{equation}
f(\textbf{v})=\frac{1}{N_{\rm esc}(2\pi\sigma_v^2)^\frac{3}{2}}\exp\left(-\frac{\textbf{v}^2}{2\sigma_v^2}\right)\Theta(v_{\rm esc}-|\textbf{v}|),
\end{equation}
where $v_{\rm esc}=544 $ kms$^{-1}$~\cite{smith:2007} is the escape velocity. The dark matter velocity dispersion $\sigma_v=\frac{v_0}{\sqrt{2}}$, where $v_0=220$ kms$^{-1}$ is the local circular speed~\cite{kerr:1986}. The normalization constant ${N_{\rm esc}}$ is given by
\begin{equation}
N_{\rm esc}={\rm erf}\left(\frac{v_{\rm esc}}{\sqrt{2}v_0}\right)-\sqrt{\frac{2}{\pi}}\frac{v_{\rm esc}}{\sigma_v}\exp\left(-\frac{v_{\rm esc}^2}{2\sigma_v^2}\right).
\end{equation}
Applying a Galilean transformation on the velocity distribution in the Galactic frame, we obtain the distribution in the laboratory frame as 
\begin{equation}
 f_{\rm lab}(\textbf{v})=f_{\rm gal}(\textbf{v}+\textbf{v}_{\rm lab}).
\end{equation}
 The velocity of the laboratory frame $\textbf{v}_{\rm lab}$ with respect to the galactic rest frame is given by
\begin{equation}
\textbf{v}_{\rm lab}=\textbf{v}_{\rm GalRot}+\textbf{v}_{\rm Solar}+\textbf{v}_{\rm Earth}+\textbf{v}_{\rm EarthRot},
\end{equation}
where $\textbf{v}_{\rm GalRot}$, $\textbf{v}_{\rm Solar}$, $\textbf{v}_{\rm Earth}$, and $\textbf{v}_{\rm EarthRot}$ are the galactic rotation, the peculiar velocity of the Solar system with respect to the galactic halo, Earth’s revolution, and Earth’s rotation velocities respectively. Even though the rotation velocity of the Earth is small compared to the other components, it plays an essential role in the daily modulation of the interaction rate, due to the effect on the direction of the dark matter velocity in the laboratory frame. The relations and algebraic calculations for converting the velocity and coordinates from the galactic frame to the laboratory frame are given in the reference~\cite{bozorgnia:2011}.
Since the direction of the laboratory velocity is towards the constellations of Cygnus~\cite{OHare:2015}, the direction of the velocity of the dark matter can be considered as the inverse of the direction of Cygnus.

\begin{figure} 
 \centerline{\includegraphics[width=9cm]{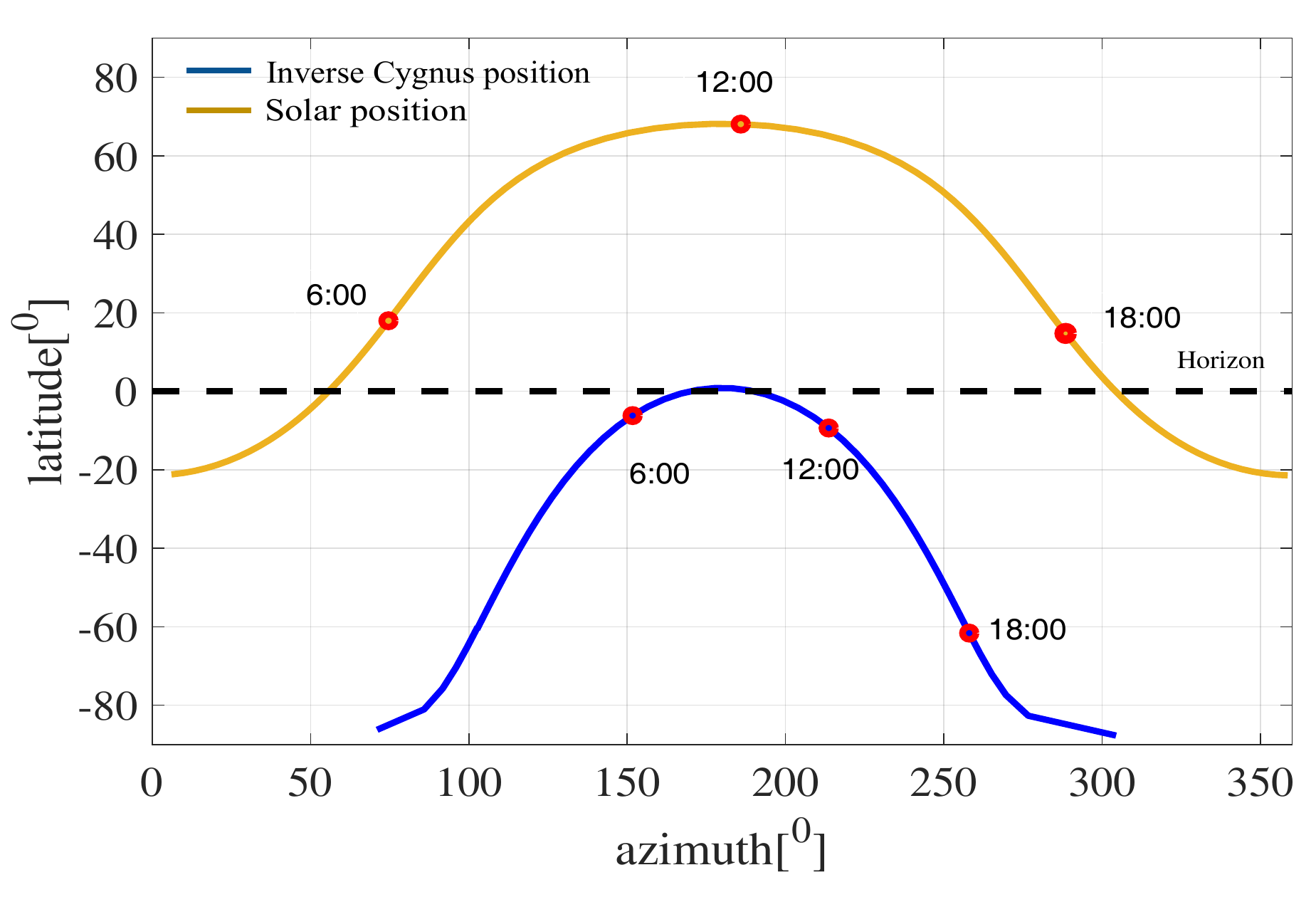}}
\caption[]{The position of the Sun (yellow) and the direction of the dark matter flux (blue) during a day in latitude 45.2° and longitude 6.7° on June 24, 2020. The red dots indicate the positions at 6, 12, and 18 o'clock. The dashed line in the middle is the horizon. }
 \label{cyg&Sun}
\end{figure}

\begin{figure} 
 \centerline{\includegraphics[width=9cm]{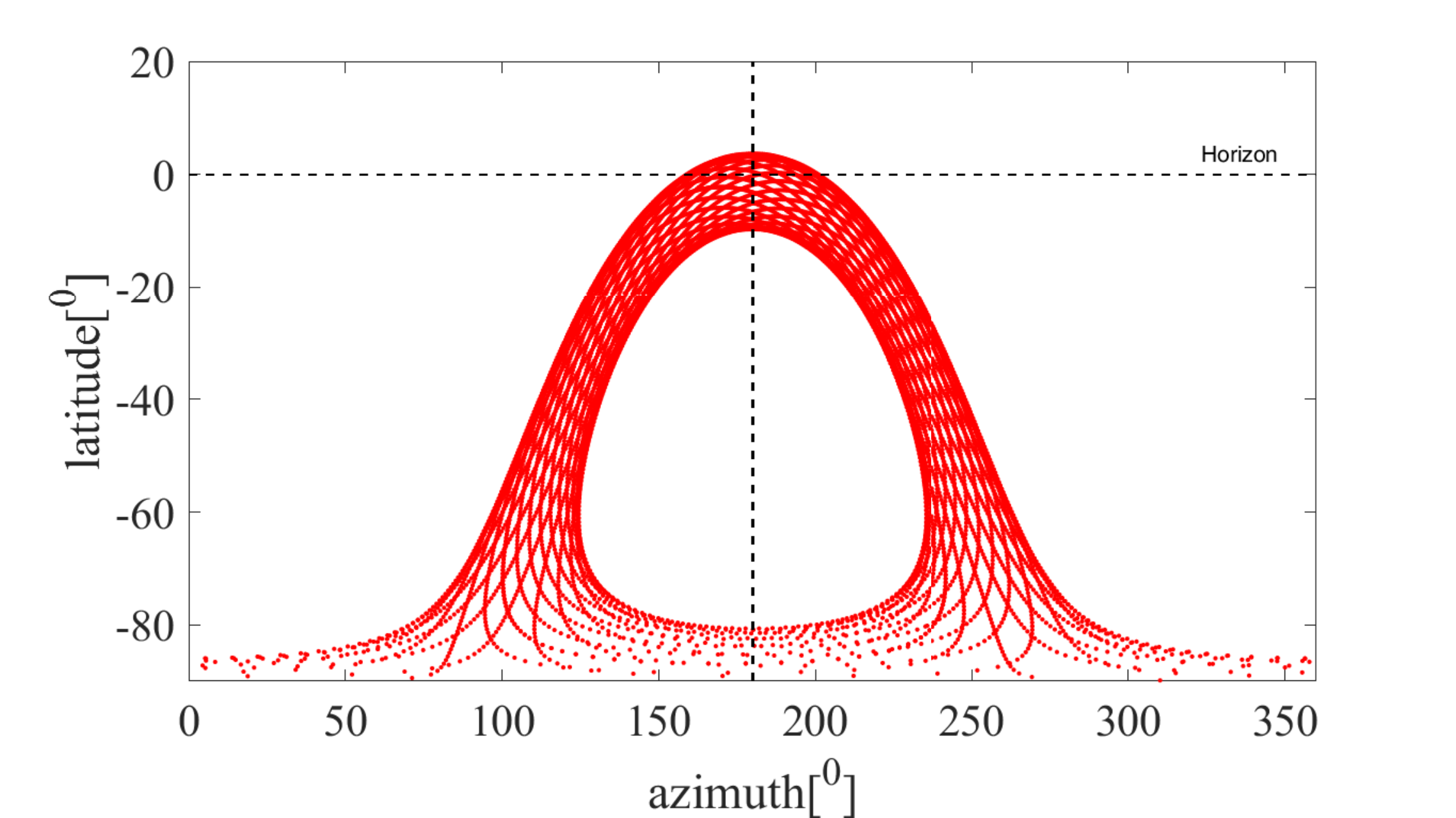}}
 \caption[]{The hourly changes in the position of the constellation Cygnus during a year for latitude 45.2$^0$ and longitude 6.7$^0$. Each red dot corresponds to one hour.}
 \label{cygnus}
\end{figure}

Fig.~\ref{cyg&Sun} shows the position of the Sun (yellow line) and the inverse of the Cygnus's position (blue line) in the sky for a laboratory located at latitude 45.2 and longitude 6.7 on June 24, 2020. The red dots indicate the position of the Sun and Cygnus at 6, 12, and 18 o'clock. 
Since the direction of the dark matter wind is from the Cygnus and the dominant low energy neutrino background is from the Sun, it is useful to notice that these directions never overlap in the figure.
Fig.~\ref{cygnus} shows the hourly changes in the Cygnus's direction relative to the laboratory's location for latitude 45.2 and longitude 6.7 in one year. Each red dot corresponds to one hour from January 1, 2020, 0:00, until December 31, 2020, 23:00. According to the Earth's rotation around the Sun, the direction of the Cygnus changes during the  year, affecting the daily modulation of the event rate.

 The dark matter scattering event rate as a function of recoil energy, direction in the laboratory frame, and time is given by \cite{Kadribasic:2017,Heikinheimo:2019,sassi:2021},
 \begin{equation}
\frac{d^3R}{dE_rd\Omega_r dt}=\frac{\rho \sigma_{m_D-n}}{4\pi m_D \mu_{m_D-n}^2\Delta t}A^2F^2(E_r) \hat{f}_{\rm SHM}(v_{\rm min},\textbf{q},t),
\end{equation}
Where $\rho=0.3$ GeV/cm$^{3}$ is the local dark matter density, $\textbf{q}$ the recoil direction in the detector, $\sigma_{m_D-n}$ is the spin-independent dark matter-nucleon cross-section, $A$ the mass number of the target in the detector, $m_D$ the dark matter particle mass, $\mu_{m_D-n}={m_D m_n}/({m_D+m_n})$ the dark matter-nucleon reduced mass and $F^2(E_r)$ the nuclear Helm form factor~\cite{duda:2007}. Finally, $\hat{f}_{\rm SHM}(v_{\rm min},\textbf{q},t)$ is the radon transform of the dark matter velocity distribution~\cite{Heikinheimo:2019}. The analytical formula of the radon transform given by~\cite{Heikinheimo:2019},
\begin{eqnarray}
\hat{f}_{\rm SHM}(v_{\rm min},\textbf{q},t)=&\frac{1}{N_{\rm esc}(2\pi\sigma_v^2)^\frac{1}{2}} \Bigl[\exp\left(-\frac{|v_{\rm min}+\textbf{q}\cdot \textbf{v}_{\rm lab}|^2}{2\sigma_v^2}\right)\nonumber \\
&-\exp(-\frac{|v_{\rm esc}|^2}{2\sigma_v^2})\Bigr],
\label{f_SHM}
\end{eqnarray}
where $v_{\rm min}=\sqrt{2m_n E_r}/{2\mu_{m_D-N}}$ is the minimum velocity required to produce a nuclear recoil with energy equal to $E_r$, where $\mu_{m_D-N}$ is the dark matter-nucleus reduced mass and $\sigma_v$ is the velocity dispersion.
Equation (6) clearly shows that when $v_{\rm min}<v_{\rm lab}$ , in the case of $\textbf{q}\cdot \textbf{v}_{\rm lab}=-v_{\rm min}$ or $\textbf{q}=- \textbf{v}_{\rm lab}$, the rate has a maximum. In other words, the relationship between the recoil direction and the laboratory motion in the galactic rest frame confirms the existence of a dark matter signal.

\begin{figure}[ht]
\centering
\includegraphics[width=3 in]{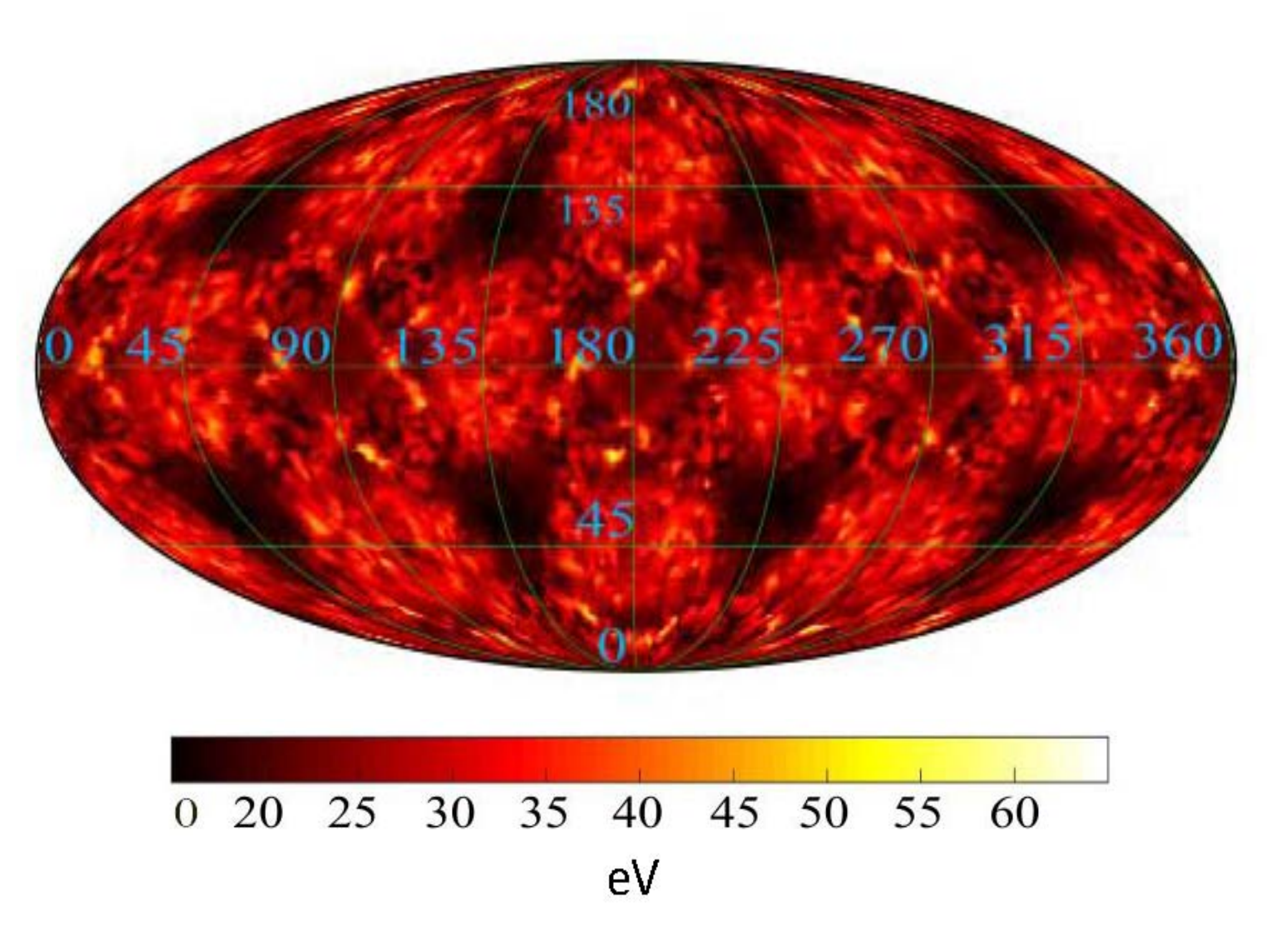}
\qquad
\caption[]{Angular distribution of the defect creation energy threshold of  silicon as a function of the recoil direction.}
\label{threshold_energy}
\end{figure}

To determine the recoil direction in a detector, we consider the axes $\hat{\textbf{x}}$, $\hat{\textbf{y}}$, and $\hat{\textbf{z}}$ as the north, east and vertical directions, respectively,
\begin{equation}
\hat{q}=s\sin(\theta)\cos(\phi)\hat{\textbf{x}}+\sin(\theta)\sin(\phi)\hat{\textbf{y}}+\cos(\theta)\hat{\textbf{z}}.
\end{equation}
In an ionization detector, the total accepted signal rate is obtained by integrating the differential rate over the recoil direction and recoil energy
\begin{equation}
R(t)=\int_{4\pi}\int_{E_{\rm th}(\theta,\phi)}^{E_r^{\rm max}}\frac{d^2R}{dE_rd\Omega_r}{dE_rd\Omega_r},
\label{eq_R}
\end{equation}
where $E_{\rm th}$ is the minimum observable recoil energy, which is here assumed to be correlated with the threshold displacement energy, and therefore depends on the recoil direction with respect to the crystal lattice~\cite{Kadribasic:2017,Heikinheimo:2019}. The threshold energies are obtained from molecular dynamics simulations~\cite{Kadribasic:2017}. The silicon threshold ranges from 17.5 to 77.5 eV.  
Fig.~\ref{threshold_energy} represents the angular distribution of the threshold displacement energy. The brighter points correspond to directions of higher threshold energy, indicating that more energy is needed to create a defect, and according to our assumption, to create a detectable ionization signal. The energy threshold surface shown in Fig.~\ref{threshold_energy} therefore determines the lower limit of the energy integral in equation~\ref{eq_R}.
\\
\section{Results}\label{sec3}
\begin{figure}[ht]
\centering
\includegraphics[width=6cm]{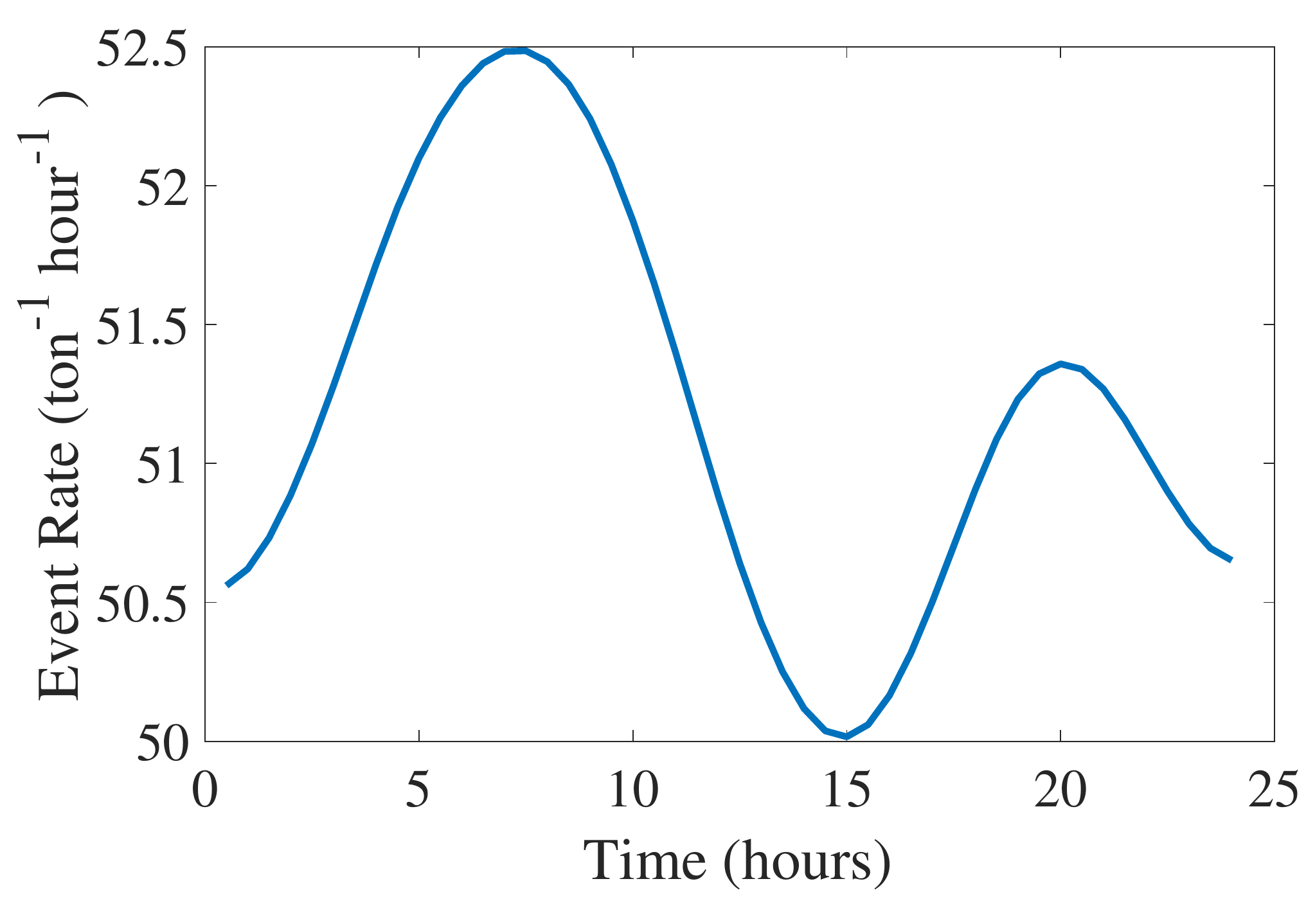}
\includegraphics[width=6cm]{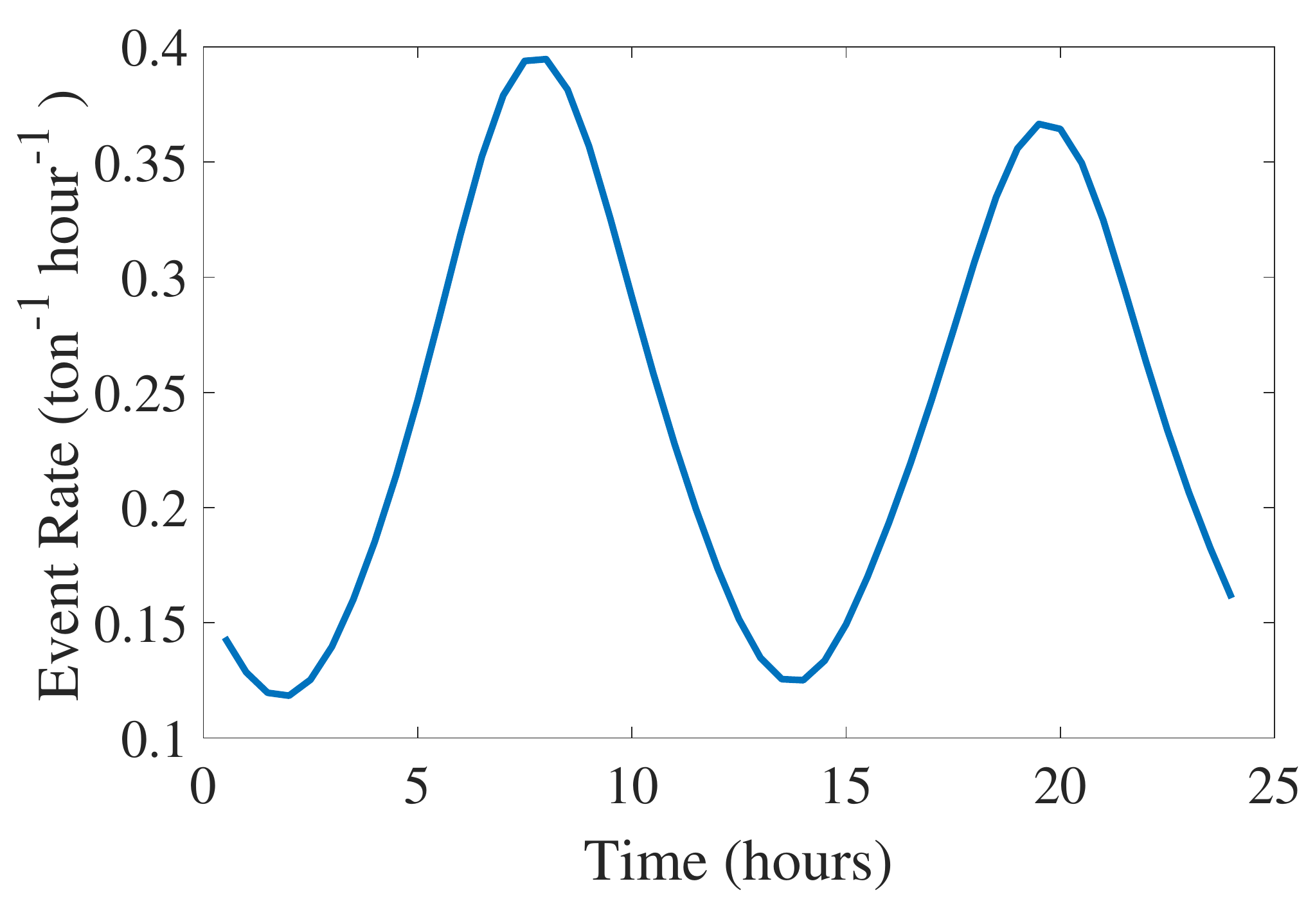}
\qquad
\caption[]{Time evolution of the event rate for a 300 MeV/c$^2$ dark matter particle with a 10$^{-39}$ cm$^2$ cross-section in interaction with silicon (top) and germanium (bottom).}
\label{Ge&Si_daily}
\end{figure}

Here, we study the time evolution of the event rate of dark matter interacting with silicon and germanium nuclei. The dark matter event rate is obtained for each time bin by averaging 24,155 and 84,936 directions for silicon and germanium, respectively. We calculate the rate by integrating over the recoil energy for each sampled direction ($\theta_i$,$\phi_i$) with the corresponding threshold energy value ($E_\text{min}$($\theta_i$,$\phi_i$)). Throughout the analysis, we take the detector mass to be one ton. Fig.~\ref{Ge&Si_daily} shows the time evolution of the event rate of a dark matter particle of mass 300 MeV/c$^2$ 
interacting with silicon (top) and germanium (bottom) 
via spin-independent cross-section 10$^{-39}$ cm$^2$
for January 1, 2020. As expected the 
event rates of both detectors oscillate due to the Earth's rotation. As seen in Fig.~\ref{Ge&Si_daily}, for the germanium detector the overall rate is much lower than for silicon. The difference in the total rate is mostly due to the larger nuclear mass in germanium, implying a larger $v_{\min}$ for a given recoil energy, and therefore cutting out a larger fraction of the DM velocity distribution.

\begin{figure}[ht]
\centering
\includegraphics[width=8cm]{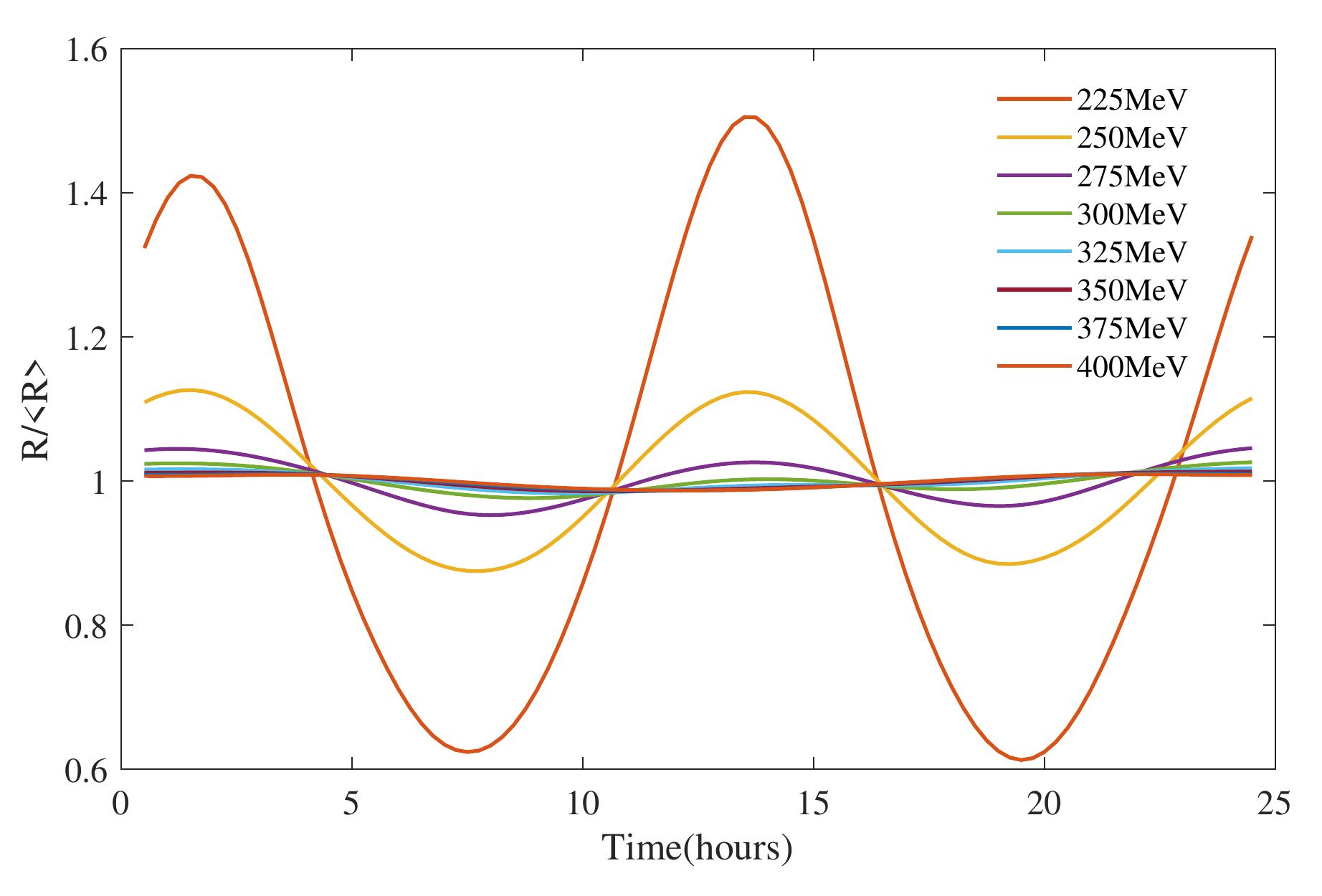}
\qquad
\caption[]{Normalized diurnal modulation of the event rate for different masses range of dark matter from 225 to 400 MeV/c$^2$ interacting with silicon nuclei.}
\label{Si_mass}
\end{figure}

The normalized diurnal modulation of the event rate for the dark matter mass range from 225 to 400 MeV/c$^2$ interacting with silicon nuclei is depicted in Fig.~\ref{Si_mass}. It was already noted in~\cite{Kadribasic:2017,Heikinheimo:2019} that the modulation amplitude increases for decreasing dark matter mass for a germanium target. Here we observe the same effect for silicon. For low-mass dark matter, the diurnal modulation effect can be utilized to discriminate dark matter from the background.

\begin{figure}[ht]
\centering
\includegraphics[width=8cm]{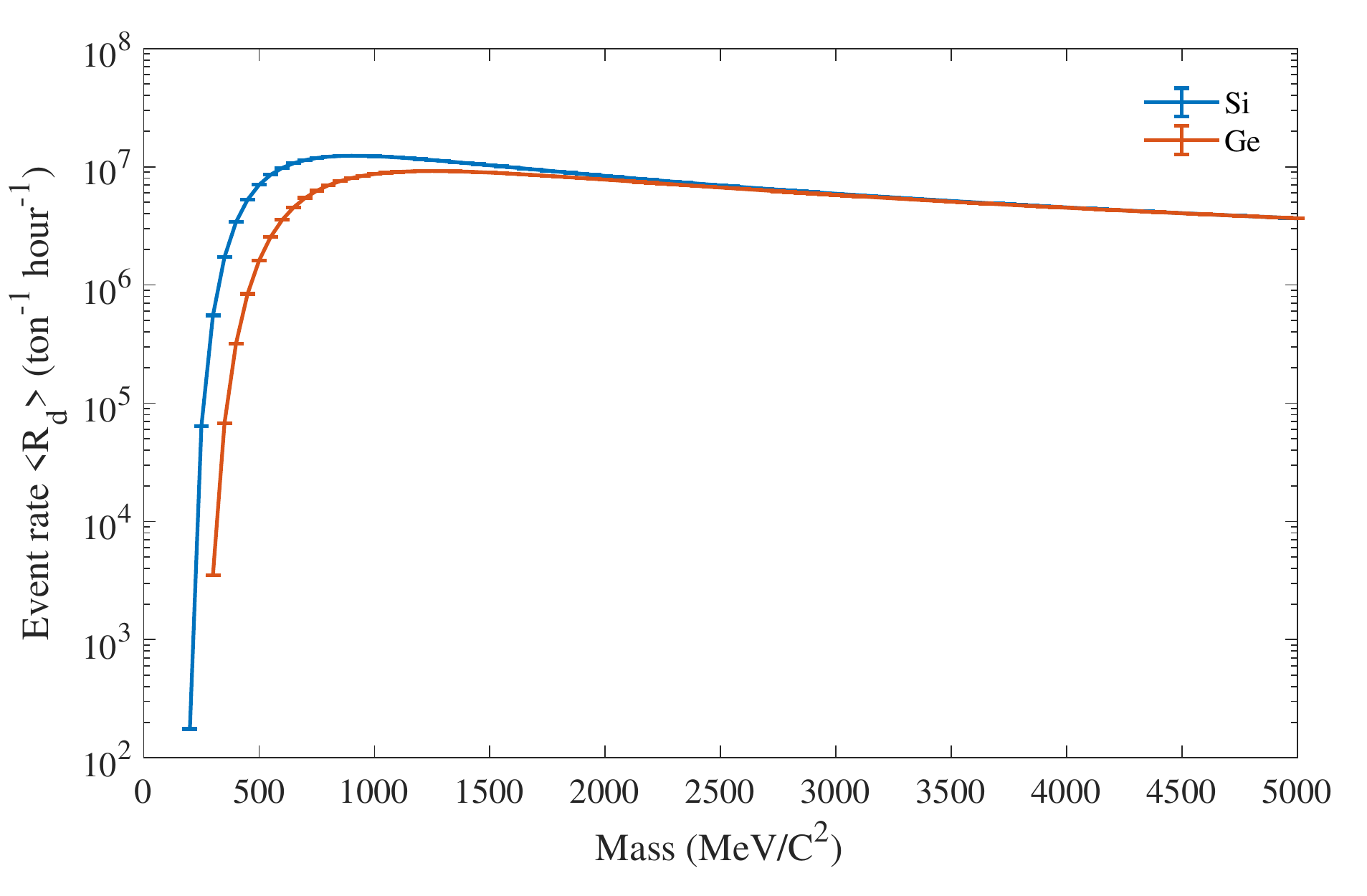}
\qquad
\caption[]{The daily average together with error bars of the event rate for the dark matter mass range from 200 MeV/c${^2}$ to 5 GeV/c$^2$ for silicon (blue line) and germanium (red line) on January 1, 2020. } 
\label{R_mass}
\end{figure}

Silicon offers a better sensitivity to low mass dark matter due to its smaller atomic number. This is evident in Fig.~\ref{R_mass}, showing the daily average of the event rate for the dark matter mass range from 200 MeV/c$^{2}$ to 5 GeV/c$^{2}$ for silicon (blue line) and germanium (red line) on January 1, 2020. 
The event rate for silicon is about 200 times larger than for germanium at low mass (300 MeV/c$^{2}$). The high event rate for silicon detectors provides more sensitivity for direct dark matter searches for low masses. For high-mass dark matter (1 GeV/c$^{2}$), the event rate for the silicon detector is about 1.5 times greater than the germanium rate. Our numerical studies show that the event rate for silicon detector approaches germanium at dark matter masses larger than 1 GeV/c$^{2}$. 

\begin{figure}[ht]
\centering
\includegraphics[width=8cm]{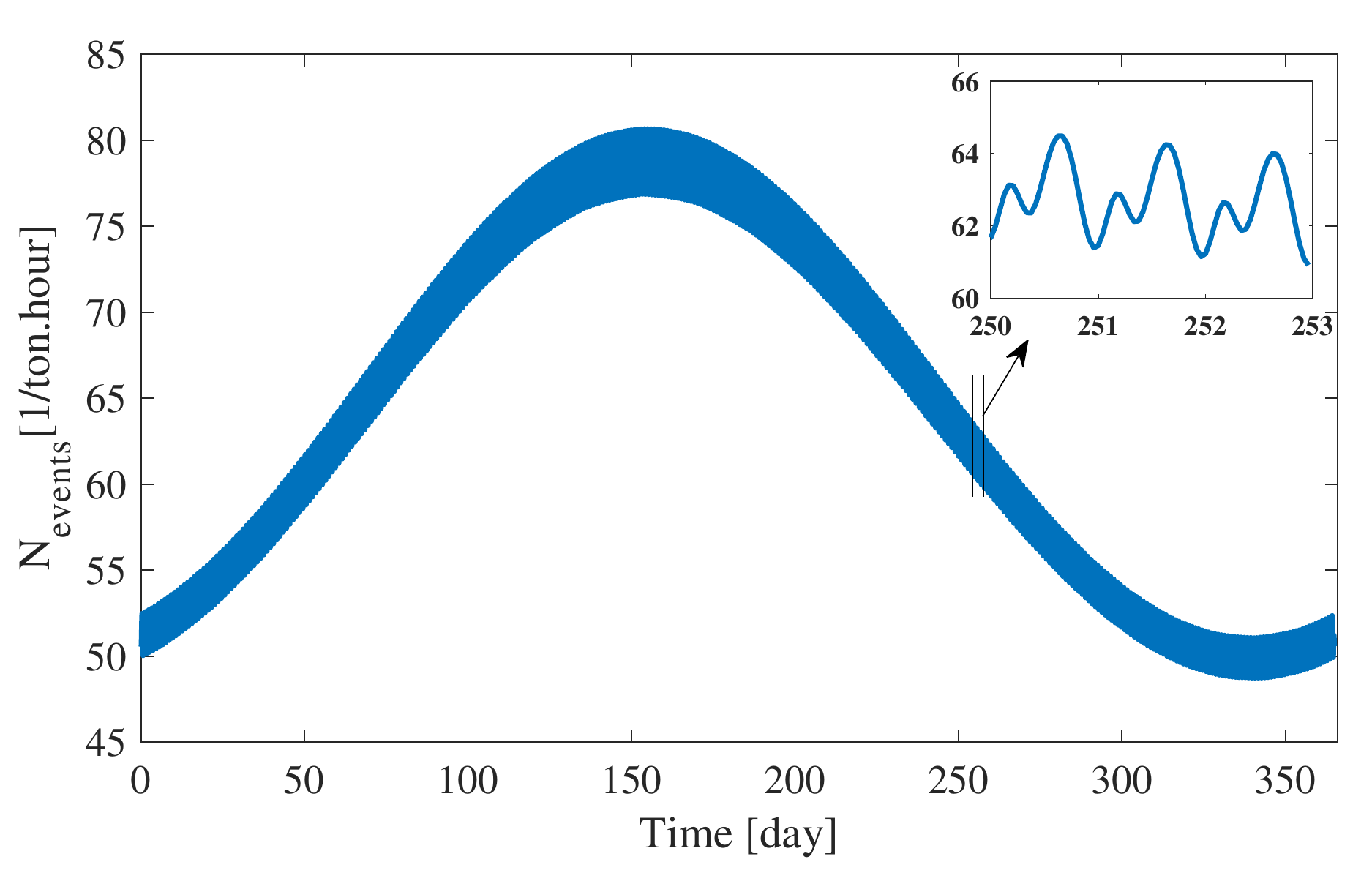}
\qquad
\caption{ The annual event rate on silicon for a 300 MeV/c$^2$ dark matter particle with a $10^{-39} cm^2$ cross-section.}
\label{Si_anual} 
\end{figure}

Fig.~\ref{Si_anual} represents the annual modulation of a dark matter particle with  mass of 300 MeV/c$^2$ and a cross-section of 10$^{-39}$ cm$^{2}$ scattering on silicon target over the year 2020.
Due to the rotation of Earth around the Sun, the orientation of the detector with respect to the inverse direction of Cygnus varies throughout the year.
Although the fluctuation reaches its maximum in June, Fig.~\ref{Si_anual} shows that the diurnal modulation remains significant for the low mass of 300 MeV/c$^2$ throughout the year. Since the direction of the neutrinos is from the Sun, the annual interaction rate of dark matter can be considered a helpful tool for distinguishing between dark matter signal and the neutrino background, as discussed in~\cite{sassi:2021}.
To further analyze the periodicity of the event rate, we applied the fast Fourier transformation (FFT) to the event rate time series (Fig.~\ref{Si_anual}). Table~\ref{tab1} represents the periods and amplitudes for silicon event rate. As shown in the table, beyond the leading annual modulation ($\sim$8760 hours), we observed the main periods of about 8 hours (three maxima per day) and 12 hours (two maxima per day).
The patterns with the period of 8 and 12 hours are due to silicon crystal orientation in laboratory coordinates (about 45-degree patterns of the angular distribution of the defect creation energy threshold for the silicon detector, Fig.~\ref{threshold_energy}). As the Earth rotates around its axis, the inverse direction of Cygnus scans various patterns of the minima (dark regions in Fig.~\ref{threshold_energy}) and maxima (bright regions in Fig.~\ref{threshold_energy}) of the threshold energy throughout the day. Depending on the orientation of the crystal frame, the path scanned by the inverse direction of Cygnus may cross two (12-hour period) and three (8-hour period) minima/maxima of the threshold surface during a daily cycle.
\begin{table}
\center
\caption{The period and amplitudes of event rate determined by the fast Fourier transformation (FFT) to the event rate time series.}
\begin{tabular}{ccc}
\\\hline
                     &  Period (hour)            & Amplitudes (Normalized) 
 \\\hline
$\mathrm{T_1}$       &   $8756\pm 53.6$          & $(225.4\pm 2.4)\times10^{-3}$      \\
$\mathrm{T_2}$       &   $4170\pm 19.1$          & $(3.8\pm 0.07)\times10^{-3}$ \\
$\mathrm{T_3}$       &   $ 24\pm 0.0005$         & $(2.8\pm 0.04)\times10^{-3}$ \\
$\mathrm{T_4}$       &   $12\pm 0.0001 $         & $(13.1\pm 0.16)\times10^{-3}$ \\
$\mathrm{T_5}$       &   $ 8\pm  0.0008$         & $(2.9\pm 0.05)\times10^{-3}$                   \\\hline
\end{tabular}
\label{tab1}
\end{table}
\section{Conclusions}
We have studied the dark matter interaction rate in silicon assuming that the ionization threshold directional dependence follows similar functional form to the threshold displacement energy simulated using molecular dynamics methods. ~\cite{Holms:2008}. 
We obtained a daily and annual modulation of the dark matter event rate for the silicon target. We found that the directional dependence of the threshold energy and the motion of the laboratory with respect to the galactic rest frame results in a modulation of the dark matter event rate in silicon, thus it provides a cosmological signature to identify dark matter from background interactions.
Therefore silicon appears as very appropriate target for identifying dark matter from backgrounds with a high event rate expected for low-mass dark matter. Using  FFT analysis, we obtained the 8 and 12 hours periodicity in simulated time series of event rates for the silicon detector that is due to the 45-degree pattern in the displacement threshold energy surface of silicon. We also observed the periods 24 hour, six months, and one year as the harmonics of the primary modes (8 and 12 hours).

\bibliographystyle{apsrev4-1}
\bibliography{apssmple}
\end{document}